\documentclass[journal,twoside,web]{IEEEtai}
\usepackage{cite}
\usepackage{amsmath,amssymb,amsfonts}
\usepackage{algorithmic}
\usepackage{graphicx}
\usepackage{booktabs}
\usepackage{multirow}
\usepackage{algorithm,algorithmic}
\usepackage{hyperref}
\hypersetup{hidelinks=true}
\usepackage{textcomp}
\usepackage{latexsym}
\usepackage{verbatim}
\let\labelindent\relax
\usepackage{enumitem}

\usepackage{fancyhdr}

\def\BibTeX{{\rm B\kern-.05em{\sc i\kern-.025em b}\kern-.08em
    T\kern-.1667em\lower.7ex\hbox{E}\kern-.125emX}}
\begin{document}
\title{Clinical Reading Comprehension with Encoder-Decoder Models Enhanced by Direct Preference Optimization}

\author{Md Sultan Al Nahian, Ramakanth Kavuluru
\thanks{Md Sultan al Nahian is with the University of Kentucky, Lexington, KY, 40508, USA (email: sa.nahian@uky.edu). }
\thanks{Ramakanth Kavuluru is with the University of Kentucky, Lexington, KY, 40508, USA (ramakanth.kavuluru@uky.edu). }}

\maketitle

\begin{abstract}
Extractive question answering over clinical text is a crucial need to help deal with the deluge of clinical text generated in hospitals. While encoder models (e.g., BERT) have been popular for this reading comprehension task, recently encoder-decoder models (e.g., T5) are on the rise. There is also the emergence of preference optimization techniques to align decoder-only LLMs with human preferences. In this paper, we combine encoder-decoder models with the direct preference optimization (DPO) method to improve over prior state of the art for the RadQA radiology question answering task by 12--15 F1 points. To the best of our knowledge, this effort is the first to show that DPO method also works for reading comprehension via novel heuristics to generate preference data without human inputs. 

\end{abstract}


\section{Introduction}
\label{sec:introduction}

Clinical text is indispensable for patient care (e.g, for tracking patient progress, care transfers) and constitutes a significantly large portion of electronic medical records (EMRs) in the U.S. As clinical notes are authored by physicians, it is also important to give them the ability to ask simple questions that can be answered from these notes rather than burdening them to peruse multiple notes for each query. From a natural language processing (NLP) perspective, this can be modeled as a machine reading comprehension (MRC) task where the answer to a question is expected to be contained as a span of text in an input document. In this paper, we achieve state of the art results for this task in radiology, with encoder-decoder language models (LMs) enhanced by recent advances in direct preference optimization (DPO). Before we proceed, we first trace the origins of DPO since it was first introduced for a very different purpose than reading comprehension. 

Since mid 2020, large language models (LLMs) have become pivotal in natural language processing (NLP), showcasing remarkable performance across a variety of tasks involving generation and domain knowledge in various fields. These models undergo an initial phase of unsupervised pretraining, where they acquire a comprehensive language representation that equips them with robust and contextual generation capabilities, which can then be transferred to perform specific downstream tasks through supervised fine-tuning \cite{dai2015semi, peters-etal-2018-deep, radfordimproving, devlin-etal-2019-bert, urvashi2019efficientlm}.
However, while supervised fine-tuning has been proven effective in enhancing model performance, aligning these models with human preferences by only this method poses significant challenges \cite{learning_to_summarize}.
The high-quality output achieved through supervised fine-tuning often poorly correlates with human judgment, as the maximum likelihood objective struggles to capture the nuances of human preferences \cite{chaganty-etal-2018-price, ondrej2017nlg}. This became quite prominent in the context of LLM chat bots where users can ask questions on objectionable or controversial topics (e.g., race, gender, violence, self-harm, and toxicity) and LLMs ought to respect social and cultural preferences when responding rather than regurgitating offensive or harmful content from pretraining corpora. 
To address this challenge, reinforcement learning from human feedback (RLHF) has recently emerged as a promising approach for aligning LLMs with human preferences\cite{ziegler2019fine, learning_to_summarize}. RLHF utilizes human feedback on the model’s output to guide its learning process, resulting in enhanced performance and better correlation with human judgment across diverse NLP tasks \cite{ouyang2022training, glaese2022improving, bai2022training}.

Ability to evaluate the output of LLMs based on human preferences is a core part of RLHF. To acquire this ability, the RLHF technique involves building a reward model or preference model from human annotated preference data. The objective of the reward model is to assess the output of the language model based on human preferences and represent it in a scalar value, which is used to optimize the language model using RL algorithms, most commonly proximal policy optimization (PPO) \cite{ppo_2017}. 
Usually the reward models are built by fine-tuning another LLM as it is expected that the reward model should have the similar language modeling capabilities to the original language model it is used to optimize.


While RLHF demonstrates impressive performance across various NLP tasks \cite{chowdhery2023palm, touvron2023llama}, it is a complex and computationally expensive process that involves training multiple models, including a supervised fine-tuned model, a reward model, and the final RLHF model. To address this complexity, Rafailov et al.~\cite{rafailov2024direct} introduced Direct Preference Optimization (DPO), which directly learns human preferences from the preference dataset to optimize language models, eliminating the need to train a reward model. This simplifies the process while maintaining the same optimization objectives as RLHF. It uses the well understood binary cross-entropy loss, without applying RL algorithms to optimize LMs.
Since it does not require training a reward model, DPO is computationally less expensive and more dynamic. 
Furthermore, compared to PPO-based RLHF, DPO focuses on both preferred and rejected data, enabling it to learn not only what to generate but also what not to generate. This flexibility allows DPO to align the language model with diverse user preferences in various NLP tasks.

Thus far DPO has been primarily used to align decoder-only LLMs with human preferences; it has not been applied to encoder-decoder models used for the MRC task (or any other information extraction task) with a likelihood maximization objective. DPO inherently aims to increase the log probability of expected outputs over rejected outputs. 
A dataset of diverse instances of correct and incorrect output pairs can provide proper signals to the model about challenging examples that a supervised fine-tuned model struggles to predict accurately.
Based on this observation, we hypothesize that DPO can be utilized to enhance the performance of a supervised fine-tuned encoder-decoder model in log-likelihood maximization.
To test this, we experiment with a recent biomedical MRC dataset, Radiology Question Answering (RadQA) \cite{soni-etal-2022-radqa}, resulting in  the following contributions and findings:
\begin{itemize}
    \item Compared with the encoder-only models used in prior efforts with RadQA, we show over 10\% F-score improvement by shifting to encoder-decoder models, achieving a new state of the art (SoTA) score. 
    \item We introduce two new methods to automatically generate paired preference data for the MRC task and use them to produce additional 1-3\% F1 gains with DPO, leading to overall gains of 12--15\% F1 points over SoTA. 
\end{itemize}

The code  used in our experiments and the preference data we created will be shared on GitHub if the paper is accepted. The original RadQA dataset is already public: \url{https://physionet.org/content/radqa/1.0.0/}.

\section{Related Works}
\subsection{Machine Reading Comprehension}

MRC is a key research area within the information extraction domain that focuses on enabling machines to extract answers from given texts. Specifically, an MRC model receives a context and a question as input and aims to accurately answer the question by reasoning over the provided context and the question itself. 
While MRC is important in and of itself, it also plays a crucial role in open ended QA where an initial retrieval model extracts relevant documents for a question from a search index. MRC is then applied to each of these documents and the answered are ranked using other heuristics. 
Prior efforts in deep learning for MRC focused on attention mechanisms, which helped models focus on relevant parts of the query and the context  \cite{seo2016bidirectional, cui2017attention, hu2019retrieve}. 
More recently, approaches using transformer-based LMs, such as BERT \cite{devlin-etal-2019-bert}, RoBERTa \cite{liu2019roberta} and XLNet\cite{yang2019xlnet} have demonstrated superior performance on this task.
These models leverage large-scale pre-training on diverse datasets followed by fine-tuning on specific MRC tasks, which significantly enhances their ability to ``understand'' and generate accurate answers.
For example, ForceReader\cite{chen-wu-2020-forcereader} is a BERT based method that addressed the attention deconcentration problem in MRC and introduced a few novel ideas including \textit{attention separate representation}, \textit{multi-mode reading}, and \textit{conditional background attention} to improve MRC. Similarly, Lu et al.~\cite{luo-etal-2020-map} proposed a novel approach that leverages BERT and BiDAF~\cite{seo2016bidirectional}, extending probability vectors to probability matrices to predict the start and end positions of the answer span more accurately.

In our approach we also used transformer-based LMs. In contrast to the previously discussed methods, we have used an encoder-decoder transformer model \cite{raffel2020exploring} as the base model and fine-tuned it by adopting the DPO method. Thus, the most closely related work to ours involves RL-based MRC methods. Although this domain is less explored compared to other deep learning approaches discussed above, several studies have applied RL techniques in question answering systems \cite{ReasoNet2017, hu2018reinforced, lee2021mrc, gharagozlou2022rlas}. 
These approaches typically design a reward function to optimize the model using RL algorithms.
However, by leveraging the DPO technique in our method, we obviate the need of a reward function for training the model.

\subsection{Reinforcement Learning from Human Feedback}
RLHF is an RL technique that optimizes models using human feedback instead of predefined reward functions. Initially explored in the context of training RL  agents~\cite{akrour2012april} where reward functions are difficult to specify, RLHF has more recently been widely used to fine-tune LLMs to better align with human preferences. This method has been successfully applied across various natural language processing (NLP) tasks, including conversational agents \cite{chatgpt}, text summarization, dialogue summarization \cite{chen-etal-2023-human}, question answering \cite{nakano2021webgpt}, and recommendation systems, where aligning the responses with human judgment is crucial.
However, RLHF involves a multi-step process that can be computationally intensive. Direct Preference Optimization (DPO) \cite{rafailov2024direct} has emerged as a more efficient alternative, aiming to achieve similar objectives with reduced computational costs. While DPO is primarily used to align language models with human judgment \cite{tunstall2023zephyr, zhao2023beyond}, our work explores its application in likelihood maximization for MRC. By applying DPO to enhance supervised fine-tuned models, we aim to improve performance by optimizing responses to match ground truth answers more closely.

\section{Background for RLHF and DPO}

Fine-tuning LLMs for downstream tasks using RLHF technique involves three main phases \cite{learning_to_summarize, bai2022constitutional}: 1. supervised fine-tuning, 2. constructing reward model, and 3. fine-tuning the language model using RL methods.

\paragraph{Supervised Fine-tuning}
This is the initial step of RLHF technique, where the language model undergoes supervised fine-tuning on downstream tasks. 
During this phase, the model is trained on specific task-related training datasets, allowing it to adapt its pre-trained knowledge to the particular downstream task. The model trained in this phase is commonly referred to as supervised fine-tuning (SFT) model, denoted as $\pi_{sft}$. 
\paragraph{Constructing Reward Model}
After training the SFT model, the next step is to develop a reward model that evaluates the SFT model's outputs based on human preferences and represent it as scalar values. This reward model can be built using pre-trained models capable of assessing outputs according to human judgment \cite{bai2022constitutional}, or by training it on human preference data collected from annotators.


To construct human preference data, multiple responses are first generated for each prompt by the SFT model, using different variants of the model or sampling methods \cite{learning_to_summarize, bai2022training}.
The collection of prompts and their generated responses are then formatted into a batch of tuples $(x, y1, y2)$, where $x$ is the prompt and $y_1$ and $y_2$ are pair of responses sampled from the set of generated responses of the prompt $x$.
Human labelers are then instructed to choose their preferred response between the two. This process creates a preference dataset consisting of tuples $(x, y_w, y_l)$, where $y_w$ represents the preferred output and $y_l$ represents the rejected output. 

From the generated preference dataset $D$, the probability distribution of human preference can be formulated as
\begin{equation}
\label{eq:prob_pref}
    p(y_w > y_l|x) = \sigma(r(x, y_w) - r(x, y_l))
\end{equation}
using Bradley-Terry model~\cite{Bradley1952RankAO} given an optimal reward model $r$, where $\sigma$ is the logistic function. 

With the preference dataset $D = \{(x^i, y_w^i, y_l^i)\}_{i=1}^N$, we parameterize the reward model $r_{\sigma}$ and optimize it by maximizing the log likelihood of the difference between the reward of preferred response and rejected response (as in Eq.~\eqref{eq:prob_pref}) and hence minimize the loss
\begin{equation}
    \label{eq:reward_loss}
    \mathcal{L}(r_\sigma) = E_{(x,y_w,y_l)\sim D}[ -\log (p(y_w > y_l|x)) ].
\end{equation}

\paragraph{Fine-tuning Using RL method}


Finally, in this step, the trained reward model $r_\sigma$ is used to provide feedback on the output of the parameterized language model $\pi_\theta$ and optimize it by the objective of maximizing the expected reward 
\begin{equation}
    \label{eq:reward_function}
    \begin{split}
    r(x, y) & = r_{\sigma}(x, y) - \beta(\log (\pi_{\theta}(y | x)) \\
    & - \log (\pi_{ref}(y | x))),
    \end{split}
\end{equation}
where $\pi_{\theta}$ denotes the policy of the language model we are optimizing and $\pi_{ref}$ is the initial SFT model. During the RL training phase, the parameters of the SFT model $\pi_{ref}$ remain fixed. $\pi_{\theta}$ is initialized with $\pi_{ref}$ and optimized using an RL algorithm, most commonly PPO \cite{ppo_2017} and other variants of actor-critic \cite{ramamurthy2023reinforcement} algorithms. The parameter $\beta$ ensures that the trained policy $\pi_{\theta}$ will not deviate significantly from the initial SFT model $\pi_{ref}$.

While RLHF is effective, it requires training a separate reward model, which makes the overall process costly. DPO eliminates the need for a reward model by directly optimizing the language model $\pi_\theta$ using the policies of both the reference model $\pi_{ref}$ and $\pi_{\theta}$ itself. The objective function of DPO is to maximize the policy difference between the preferred output $y_w$ and the rejected output $y_l$ as in
\begin{equation}
    \begin{split}
    \mathcal{L}_{DPO}(\pi_\theta; \pi_{ref} ) = - E_{(x,y_w,y_l)\sim D} \\
    \left[\log \sigma(\beta \log \frac{\pi_{\theta}(y_w | x)}{\pi_{ref}(y_w|x)} 
    - \beta \log \frac{\pi_{\theta}(y_l | x)}{\pi_{ref}(y_l|x)}) \right].
    \end{split}
    \label{eq:dpo_loss}
\end{equation}

\begin{figure}[t]
\centering{
  \includegraphics[width=1.0\columnwidth]{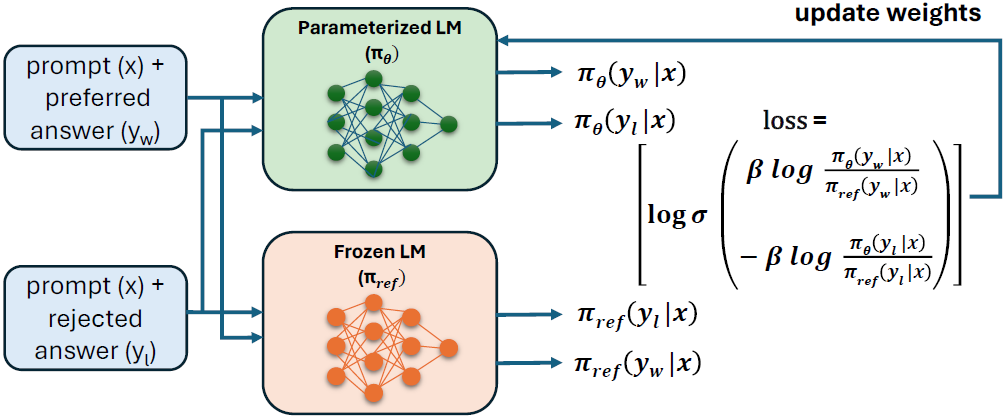} 
  
  \caption {Pipeline of fine-tuning the language model using DPO. $\pi_{\theta}$ is the language model we want to fine-tune, and $\pi_{ref}$ is the reference model, which is kept frozen during the fine-tuning process. Both models are initialized with the SFT model.}
        \label{fig:dpo}}

\end{figure}

\section{Methods}
We choose the encoder-decoder model T5~\cite{raffel2020exploring} by Google as the backbone of our main method as opposed to the BERT based baselines reported earlier~\cite{soni-etal-2022-radqa}. We also experimented with the Flan versions~\cite{longpre2023flan} of the T5 model that have been \textit{instruction tuned} on a variety of NLP datasets and tasks. 
Our DPO-based preference optimization consists of two phases: (1) training a supervised fine-tuned T5 model and (2) optimizing it using DPO.

\subsection{Training Supervised Fine-tuned (SFT) Model}

In this step, we trained an initial model for MRC using the supervised fine-tuning approach with the original training data, which we refer to as the SFT model. We model MRC as a text to text task and opted to use a seq-2-seq model for training the SFT model. The model's input is the tokenized vectors of the concatenated context and question and the output is the answer span from the context or ``no answer'' if the answer is not available in the context. We formatted the input sequence before tokenization as follows: ``context: the text of the context \textless SEP\textgreater   question: text of the question.''

\subsection{Optimizing Using DPO}
After training the SFT model,  we fine-tuned it further using the DPO method. To do this, a preference dataset is required which consists of tuples $(x, y_w, y_l)$, where $x$ is a prompt and $y_w$ and $y_l$ are the preferred and rejected responses for the prompt $x$, respectively. In standard RLHF/DPO techniques, the preference dataset is usually constructed using human annotators. For each input sequence, multiple outputs are generated by the initial SFT model and human annotators are asked to rate them as preferred or rejected outputs. 
In contrast to the standard DPO, here we constructed the preference dataset automatically without human interventions. Our approaches to create the preference dataset are discussed in  Section \ref{section:preference_data}. 

After generating the preference dataset, we applied  DPO to optimize the SFT models. The DPO architecture employs two models simultaneously for fine-tuning purposes: one serves as a reference model ($\pi_{ref}$), while the other is the active model, $\pi_{\theta}$, which is being optimized. Both models are initialized with the SFT model trained in the previous step. The weights of the reference model ($\pi_{ref}$) are kept frozen throughout the training process, while the weights of the model $\pi_{\theta}$ are updated using the DPO loss (Eq.~\eqref{eq:dpo_loss}). 
The reference model ensures that fine-tuning does not cause the policy of the model $\pi_{\theta}$ to deviate significantly from the initial SFT model. While the DPO loss aims to increase the difference between the policies for the preferred and rejected outputs, it also aims to minimize the difference between the policies of the SFT and the active model $\pi_{\theta}$.
Both models receive input in the form of the tuple $(x, y_w, y_l)$. In our study, the prompt $x$ consists of the concatenated string of the context and question, $y_w$ is the correct answer span and $y_l$ corresponds to one of the incorrect answers for the question, given the context. Given the prompt, both models provide the probability distribution of the tokens of the preferred and rejected answers, which are used to compute the loss and update the weights of the active model $\pi_{\theta}$. Figure \ref{fig:dpo} depicts the process of DPO more elaborately.

\begin{figure}[t]
\centering{
  \includegraphics[width=0.95\columnwidth]{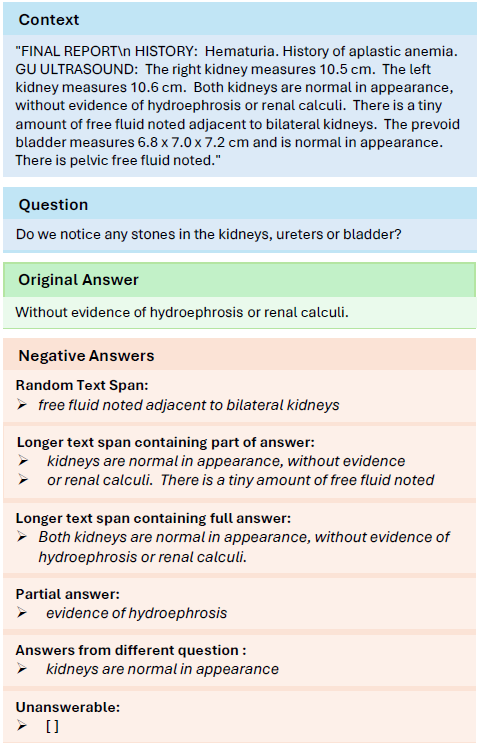} 
  \caption {Examples of negative  (rejected) outputs created by rules.}
        \label{fig:negative_example}}

\end{figure}

\section{Datasets}
We need two datasets to build the models in the two phases of our method. 
The first dataset is the original RadQA dataset, which was used for training and validating the SFT model. The second dataset, a preference dataset created from RadQA, was used for further tuning of the SFT model via DPO.
\subsection{RadQA}
RadQA\cite{soni-etal-2022-radqa} is an MRC dataset created from radiology reports from the MIMIC III dataset~\cite{johnson2016mimic}. It comprises 6148 unique question-answer pairs sourced from 1009 radiology reports of 100 patients. The dataset was split at the patient level into training, development, and testing sets, with an 8:1:1 ratio, respectively. This resulted in 4878 questions in the training set, 863 questions in the development set, and 894 questions in the test set.
We used the original format of training data of RadQA exclusively to train the SFT model, while the development and test data were used for evaluating both the SFT and DPO models to assess the effectiveness of our proposed method. 

\subsection{Preference Dataset}
\label{section:preference_data}


Preference data is the main element for optimizing a language model through DPO. This  consists of tuples that include examples of preferred and rejected outputs for a given prompt.
Although preference data is typically collected from human annotators, we automatically generated it, eliminating the need for manual annotation. We used the original training corpus of RadQA for this purpose. Specifically, each prompt was formed by concatenating the context and question from the RadQA training dataset, separated by a special token. The preferred output is the original gold answer span provided in the dataset. To generate the corresponding rejected output, we propose two automated approaches: a model-based approach and a rule-based approach.

\subsubsection{Model based approach}
In this approach, we used the SFT model itself to generate negative examples. The process began by training a model on 50\% of the RadQA training data and then using it to predict answers for the entire training dataset, including the data it was trained on. The rationale behind training on half of the data was to equip the model with sufficient knowledge for effective performance. Thus, mistakes made during these predictions indicate the types of examples the model needs to focus on to improve its performance. Testing the model on both seen and unseen data helps identifying specific examples that remain challenging despite prior exposure. 
Our intuition behind this design is that by using the model's own incorrect predictions, we can better identify the types of examples where it struggles. These incorrect predictions highlight situations where the model needs improvement, making them valuable for training. Additionally, since the model is also tested on examples it was trained on, any errors it makes on these familiar examples indicate that they are particularly challenging. By focusing on these hard examples, we aim to improve the model's overall performance.

\begin{table}[h]
    \centering
    \renewcommand{\arraystretch}{1.2}
    \begin{tabular}{lccc}
        \toprule
        \multirow{2}{*}{Dataset} & \multicolumn{3}{c}{F1 Threshold} \\
        \cmidrule{2-4} 
        {} & 0.9 & 0.7 & 0.5 \\
        \midrule
        Preference - Model-based-T5 & 3280 & 2865 & 2354\\
        Preference - Model-based-Flan-T5 & 3089 & 2533 & 2036 \\
        Preference - Rule-based & 3716 & 3501 & 3332 \\

        \bottomrule
    \end{tabular}
\caption{Number of instances in the preference dataset created by each method applying different F1 threshold values.}
\label{table:dataset}
\end{table}
We identified all instances where the model generated incorrect answers. For each prompt and question pair where the model's prediction differed from the original answer, the incorrect prediction was recorded as the rejected output in our preference dataset. To refine the preference dataset, we filtered these incorrect answers based on their F1 scores. The F1 score was calculated by comparing word-level matches between each incorrect answer and its corresponding original answer.
To filter the incorrect predictions, we applied three different thresholds for the F1 score: 0.9, 0.7, and 0.5. If the F1 score between the original and the predicted answer was less than the chosen threshold, the predicted answer was selected as the rejected output.
To ensure comprehensive coverage, we repeated this process by training another model on the remaining 50\% of the training data. This model was then used again to predict answers for the entire dataset, allowing us to identify additional incorrect predictions. We used two variants of SFT models (T5-3B and Flan-T5-3B) to create the negative examples. The total number of instances created by this process is shown in Table \ref{table:dataset}.

By iteratively training on different halves of the dataset and collecting incorrect predictions, we effectively created a robust set of negative examples without the need for manual annotation. This automated generation of preference data not only streamlined our process but also ensured a diverse range of negative examples, enhancing the overall quality of our preference dataset. Our assumption is that DPO will help the model improve on these challenging examples, thereby enhancing overall performance.

\subsubsection{Rule based approach}
We generated negative examples from the training data by applying a set of predefined rules. These rules were formulated based on experimental findings regarding the types of errors that SFT model typically makes. For each tuple (context, question, gold answer) in the training data, we generated a number of incorrect answers applying the following rules (also shown in Figure~\ref{fig:negative_example}):
\begin{itemize}[leftmargin=*]
    \item \textit{Random text span:} Select a random span from the context that does not contain any part of the gold answer.
    \item \textit{Text span containing part of the gold answer:} Here, a text span from the context that includes a part of the original answer is randomly chosen. This partial inclusion can occur in two ways: 1) choosing a segment starting a few words before the left side of the gold answer and continuing until it includes a partial span from the gold answer, or 2) selecting a partial segment from the right side of the answer and including a few words after the answer text. The lengths of these segments are chosen randomly (see Figure \ref{fig:negative_example}).
    
    \item \textit{Longer answer:} This entails a text span that includes the entire gold answer as a part of it with $\geq 1$ additional tokens.
    \item \textit{Partial answer only:} This involves selecting a smaller  segment (strict substring) from the original answer.
    \item \textit{Answers of a different question:} 
Here, an answer text from another question in the same context is chosen, provided it is not the same as the original gold answer or a part of it. For example in Figure~\ref{fig:negative_example}, ``kidneys are normal in appearance" is an answer to a different question for the same context, but  is not part of the ground truth answer.
    \item \textit{No answer:} In this approach, we used empty string in place of the gold answers to create negative examples. For questions without available answers, we chose responses from other questions within the same context as negative examples. If there were no other questions within the same context that provided answers, we randomly selected a span from the context as the negative answer.
\end{itemize}
Following these rules provided us with a large number of examples of rejected answers for each (context, question, gold answer) tuple. From each set of rejected answers, we randomly chose a few examples to create the preference data. We did not include the entire set of rejected answers for generating the preference data to prevent the dataset from becoming intractably large. Finally, we included 4000 instances and further filtered them by applying F1 threshold (see Table \ref{table:dataset}).

\section{Experimental Setup}

\subsection{Baselines}

We compared our base T5-based SFT models with the BERT-based models from Soni et al.~\cite{soni-etal-2022-radqa}, which offered SoTA results on the RadQA dataset. Thus, we selected all of their BERT-MIMIC based models as our baselines. We also compared our DPO-based method with the T5 SFT models to assess the effectiveness of applying DPO on an already high-performing fine-tuned model.

\subsection{Evaluation Metrics}
To evaluate our proposed method, we used the standard MRC metrics: Exact match (EM) and F1-Score. Exact Match is a strict metric that compares the predicted answer with the exact ground truth answer, ensuring they are identical. The F1-Score, on the other hand, is calculated by taking word-level matches between the predicted and ground truth answers. To maintain consistency and comparability in our evaluation, we used the evaluation code from SQuAD \cite{rajpurkar2016squad}.
\subsection{Network Parameters}


The network parameters for each model in our experiments were chosen through hyperparameter tuning. We used the validation F1 score as an evaluation metric to select the optimal values of these parameters. 
For training both the SFT and DPO models, we employed the Adam optimizer. The learning rate for the SFT model was set to $5e^{-5}$, and for the DPO model, it was $5e^{-7}$. The weight decay was set to 0.01 for both models. The batch size was 16 for T5-Large models; however, to accommodate the 3 billion parameter models in memory, we used a batch size of 2 with gradient accumulation steps of 8. The maximum prompt length was set to 768, and the target length was 128. Early stopping was applied during the training of both the SFT and DPO models, by using the validation F1 score to select the best models.

\section{Results}


Table \ref{table:main_results} presents the main results of our experiments, comparing the performance of BERT baselines,  the T5-based supervised fine-tuned (SFT) models, and the DPO based models. The results are evaluated on the development and test sets of the RadQA dataset.

\begin{table*}[h]
    \centering
    \begin{tabular}{lllllll}
        \toprule
        \multirow{2}{*}{Model Type} & {\multirow{2}{*}{Models}} & \multicolumn{2}{c}{Dev} & \multicolumn{2}{c}{Test}\\
        \cmidrule{3-4} \cmidrule{5-6} 
        {} & {} & EM & F1 & EM  & F1 \\
        \midrule
        \multirow{4}{*}{\parbox{3.0cm}{Baseline (BERT-MIMIC) \\ (340M)}} & RadQA  & 48.05 & 65.85  & 45.73 & 60.08 \\
        {} & emrQA-RadQA  & 50.65 & 67.97 & 47.71 & 61.60 \\
        {} & SQuAD-RadQA  & 52.28 & 69.42  & \textbf{49.39} & \textbf{63.55} \\
        {} & SQuAD-emrQA-RadQA  & 53.26 & 67.79  & \underline{48.32} & \underline{62.29} \\

        \midrule
         \multirow{4}{*}{T5-large (770M)} & {SFT} & 47.86 & 66.22 & 49.89 & 71.10 \\
         {} & {DPO-MB} & 47.74 & 66.25 & \textbf{51.34} & \textbf{71.62} \\
         {} & {DPO-RB} & 48.20 & 66.59 & \underline{51.00} & \underline{71.36}\\
         {} & {DPO-MRB} & 47.80 & 66.10 & 50.11 & 71.20\\
         
         \midrule
        \multirow{4}{*}{T5-3B} & {SFT} & 49.83 & 68.59 & 51.68 & 72.29 \\
         {} & {DPO-MB} & 51.10 & 70.45 & \underline{52.46} & \underline{74.29} \\
         {} & {DPO-RB} & 50.87 & 70.26 & \textbf{52.57} & 74.03\\
         {} & {DPO-MRB} & 50.40 & 70.13 & 52.01 & \textbf{75.18}\\
         
         \midrule
                  \multirow{4}{*}{Flan-T5-3B} & {SFT} & 54.35 & 72.62 & \underline{55.93} & 76.38 \\
         {} & {DPO-MB} & 53.77 & 73.68 & 55.15 & \textbf{77.48} \\
         {} & {DPO-RB} & 52.49 & 72.55 & \textbf{56.15} & 77.40\\
         {} & {DPO-MRB} & 53.42 & 73.51 & 55.70 & \underline{77.41}\\
         
        \bottomrule
    \end{tabular}

    \caption{Model performances on the RadQA development and test sets compared with the RadQA BERT-MIMIC model variants.}
    \label{table:main_results}
\end{table*}

The SFT model type includes three T5 variants (T5-large, T5-3B, and Flan-T5-3B) trained on the RadQA training data. From  Table~\ref{table:main_results}, we can see that all the T5 variants outperform the baseline RadQA models on the test set, with Flan-T5-3B also performing better on the dev set.
Specifically, the SFT Flan-T5-3B achieves an F1 score of 76.38 and an exact match (EM) score of 55.93 on the test set, marking improvements of 13 points in F1 score and 6.5 points in EM over the best baseline model.
Although the three variants of BERT-MIMIC were trained on additional datasets (SQuAD and emrQA) along with RadQA, the T5 models still outperformed them, establishing a strong baseline for our DPO-based method. It is important to keep in mind, however, that our T5 models have more capacity than the 340M BERT based models used by the RadQA paper. 

The DPO-based methods include three groups of models: DPO-Model Based (DPO-MB), trained on model-based preference data; DPO-Rule Based (DPO-RB), trained on rule-based preference data; and DPO-Model \& Rule Based (DPO-MRB), trained on a combined dataset of model-based and rule-based preference data. For all the models, we selected the preference data generated by 0.9 F1 threshold. Additionally, for training the DPO-MB models, we used the model specific preference data. For instance, we applied model-based-T5 preference data for the T5 models and model-based-Flan-T5 preference data for the Flan-T5 based DPO models.
From Table \ref{table:main_results} we can see that both model and rule-based DPO models improved the performance of the corresponding SFT models. Although the T5-large SFT model did not see a significant improvement, the T5-3B and Flan-T5-3B improved their corresponding SFT models nontrivially,  both in DPO-MB and DPO-RB settings. For instance, the F1 score of the DPO-MB T5-3B is 74.29, a 2-point improvement over its SFT counterpart and an 11-point increase compared to the best performing baseline model, BERT-MIMIC-SQuAD-RadQA, on the test F1 score. The combined dataset further improved the test F1 score of the T5-3B model by 1\%, but it did not enhance the other variants, indicating saturation in the performance of the models. 



 \section{Discussion}

Our experimental results demonstrate that further fine-tuning an SFT model through DPO can enhance its performance between 1--3\% F1 points. This is particularly important because these SFT models have already been optimized using the full training dataset, making further improvements challenging. From our experiments, we found several factors that influence the performance of the models trained with DPO, including the size of the SFT models, the method used to create negative examples in the preference data, the types of examples included, and the quantity of preference data. In this section, we provide a detailed discussion on the observed performance improvements using DPO and the factors influencing these improvements.


\subsection{Size of the Model}
From our experiments, we notice that a smaller model is less likely to benefit from additional fine-tuning with DPO. However, when we used larger models, notable improvements were observed. For instance, both DPO T5-3B and Flan-T5-3B increased the test F1 score of their corresponding SFT models by 1-3\%. This indicates the ability of larger encoder-decoder models to capture signals from examples of preferred and rejected outputs. However, among 3B models, the improvement is much better in the non-Flan model. Since the Flan model is instruction tuned on hundreds of datasets, its SFT performance (76.38 F1) is already over 1\% better than the best DPO model of its non-Flan counterpart.

\subsection{Model-based vs Rule-based Preference Data}
 While DPO-MB and DPO-RB both   enhanced the performance of the SFT models, our experiments showed that the model-based approach yielded comparatively better results than the rule-based approach.
One potential reason for this could be the nature of the negative examples generated by each method. Rule-based examples are created using predefined rules. Although these rules are designed to generate plausible negative examples, they may not always reflect the same distribution as the original RadQA dataset. This can lead to less effective training, as the model might not encounter a representative range of challenging examples during the DPO training.
In contrast, the model-based approach derives negative examples from the predictions of the SFT model itself. These examples are intrinsically linked to the specific weaknesses of the model. By focusing on these model-specific errors, the preference data reflects the instance spaces where the model is prone to generate incorrect outputs. Consequently, this approach may offer more targeted training, enabling the model to learn from its mistakes and improve its performance. However, one limitation of this method is that each new model requires the creation of a new preference dataset, as each model has different weaknesses and strengths. In contrast, the training examples created by the rule-based approach are model-agnostic.

\begin{figure}[t]
\centering{
  \includegraphics[width=0.9\columnwidth]{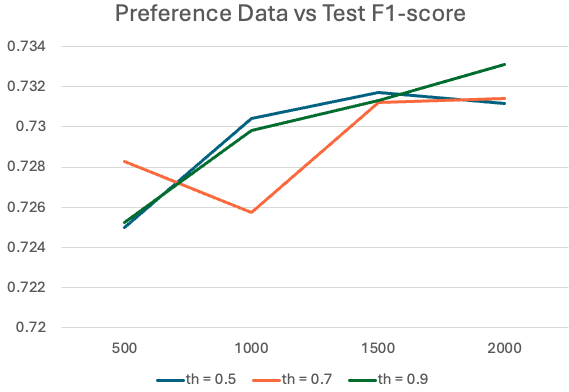} 
  \caption{Performance comparison of DPO-T5-3b model with varying training examples and preference datasets generated using different thresholds. X-axis plots the number of training examples, Y-axis is the F1 score, and the line colors represent different preference datasets created by applying three different f1 threshold.}
  
        \label{fig:preference_f1}}

\end{figure}

\subsection{Diversity of Training Instances}
Filtering the preference data based on different F1-score thresholds also influences the performance of DPO. Negative examples with higher F1 scores tend to be closer to the ground truth answers, while those with lower F1 scores present more dissimilarity with gold spans. Incorporating a broader range of negative examples from both ends of the F1-score spectrum provides a diverse and more informative training set for the model.
A higher F1-score threshold facilitates a mix of examples that are both similar and dissimilar to the ground truth answers, offering a wide variety of training data. Conversely, a lower threshold focuses only on the examples that are very different from the ground truth, excluding those that are more similar. Therefore, preference data created using higher thresholds may enable the model to learn from a diverse set of examples, which can enhance its generalization and performance.
Our experiments also support this hypothesis. Figure \ref{fig:preference_f1} illustrates the test F1-scores of DPO-T5-3B models trained with preference data filtered at different thresholds. The results show that the model trained with a threshold of 0.9 outperforms those trained with lower threshold data, demonstrating the benefits of using a more diverse set of training examples.

\subsection{Number of Training Instances}
In addition to the diversity of training examples, the number of training examples also impacts the performance of DPO based models. We  fine-tuned the DPO-T5-3B with different numbers of training examples (500, 1000, 1500 and 2000) for each filtering threshold. As shown in Figure~\ref{fig:dpo}, an increase in the number of training examples generally leads to an increase in the test F1 score across all the thresholds.

\subsection{Other Variants of DPO}
\begin{table}[h]
    \centering
        \renewcommand{\arraystretch}{1.2}
    \begin{tabular}{ p{0.9cm} p{0.51cm} p{.51cm} p{.51cm}  p{.51cm}  p{0.51cm} p{0.51cm} p{0.51cm} p{0.51cm}}
        \toprule
        \multirow{3}{*}{Loss} & \multicolumn{4}{c}{T5-3B} & \multicolumn{4}{c}{Flan-T5-3B} \\
        \cmidrule{2-5} \cmidrule{6-9} 
        {} & \multicolumn{2}{c}{Dev} & \multicolumn{2}{c}{Test} & \multicolumn{2}{c}{Dev} & \multicolumn{2}{c}{Test} \\
        \cmidrule{2-5} \cmidrule{6-9} 
        {} & EM & F1 & EM & F1 & EM & F1 & EM & F1 \\
        \cmidrule{2-5} \cmidrule{6-9} 
        DPO & \textbf{51.10} & \textbf{70.45} & \textbf{52.46} & 74.29 & 53.77 & 73.68 & 55.15 & \textbf{77.48}\\
        IPO & 50.64 & 69.41 & 51.57 & 73.41 & 53.53& 73.06&53.36 &76.79\\
        RSO & 49.83 & 69.55 & 50.90 & \textbf{74.31} & 53.88& 73.50& \textbf{55.48} &77.24\\
        KTO & 47.74 & 68.21 & 51.12 & 74.24 & \textbf{54.11} & \textbf{73.76} & 53.36 & 77.20\\

        \bottomrule
    \end{tabular}
    \caption{Results on the variants of DPO.}
    \label{table:dpo-variants}
\end{table}
DPO has evolved into several variants, each with different loss functions designed to address specific issues. For instance, Identity Preference Optimization (IPO) was developed to mitigate the overfitting problem identified in DPO, by introducing a new loss function. We trained our model using three different DPO variants: Identity Preference Optimization (IPO) \cite{azar2024general}, Kahneman-Tversky Optimization (KTO) \cite{ethayarajh2024ktomodelalignmentprospect}, and Statistical Rejection Sampling Optimization (RSO) \cite{liustatistical}. The corresponding experimental results shown in  Table \ref{table:dpo-variants} indicate that the original DPO outperforms other preference optimization techniques.

\section{Conclusion}
In this paper, we proposed an approach that combines encoder-decoder models with DPO based optimization to achieve new SoTA performances on the MRC task for radiology using the RadQA dataset. Our study shows that encoder-decoder models, although computationally expensive due to large model capacities, can offer substantial gains in performance (by over 10\% in F1 scores). Originally introduced for aligning LLMs with human preferences, our study demonstrated that DPO methods can also be effectively used for likelihood maximization for MRC tasks and can lead to further gains of up to 3\% beyond the encoder-decoder based gains. By focusing on challenging examples (the model-based preference data setup), DPO can further improve large models already fully trained. 

While effective, one key challenge in fine-tuning models using DPO is that its performance is highly dependent on the quality of the preference data. Collecting high-quality examples of preferred and rejected outputs is crucial for maximizing the model's performance through DPO. In this work, we introduced two techniques to generate preference data for the MRC task, which can be adopted in other tasks as well. In future, we will explore the applicability of our approach in other  information extraction tasks such as named entity recognition and relation extraction. 

\vspace{-2mm}
\bibliography{main}

\begin{thebibliography}{10}
\providecommand{\url}[1]{#1}
\csname url@samestyle\endcsname
\providecommand{\newblock}{\relax}
\providecommand{\bibinfo}[2]{#2}
\providecommand{\BIBentrySTDinterwordspacing}{\spaceskip=0pt\relax}
\providecommand{\BIBentryALTinterwordstretchfactor}{4}
\providecommand{\BIBentryALTinterwordspacing}{\spaceskip=\fontdimen2\font plus
\BIBentryALTinterwordstretchfactor\fontdimen3\font minus \fontdimen4\font\relax}
\providecommand{\BIBforeignlanguage}[2]{{%
\expandafter\ifx\csname l@#1\endcsname\relax
\typeout{** WARNING: IEEEtran.bst: No hyphenation pattern has been}%
\typeout{** loaded for the language `#1'. Using the pattern for}%
\typeout{** the default language instead.}%
\else
\language=\csname l@#1\endcsname
\fi
#2}}
\providecommand{\BIBdecl}{\relax}
\BIBdecl

\bibitem{dai2015semi}
A.~M. Dai and Q.~V. Le, ``Semi-supervised sequence learning,'' in \emph{Advances in Neural Information Processing Systems}, C.~Cortes, N.~Lawrence, D.~Lee, M.~Sugiyama, and R.~Garnett, Eds., vol.~28.\hskip 1em plus 0.5em minus 0.4em\relax Curran Associates, Inc., 2015.

\bibitem{peters-etal-2018-deep}
\BIBentryALTinterwordspacing
M.~E. Peters, M.~Neumann, M.~Iyyer, M.~Gardner, C.~Clark, K.~Lee, and L.~Zettlemoyer, ``Deep contextualized word representations,'' in \emph{Proceedings of the 2018 Conference of the North {A}merican Chapter of the Association for Computational Linguistics: Human Language Technologies, Volume 1 (Long Papers)}, M.~Walker, H.~Ji, and A.~Stent, Eds.\hskip 1em plus 0.5em minus 0.4em\relax New Orleans, Louisiana: Association for Computational Linguistics, Jun. 2018, pp. 2227--2237. [Online]. Available: \url{https://aclanthology.org/N18-1202}
\BIBentrySTDinterwordspacing

\bibitem{radfordimproving}
A.~Radford, K.~Narasimhan, T.~Salimans, and I.~Sutskever, ``Improving language understanding by generative pre-training.''

\bibitem{devlin-etal-2019-bert}
J.~Devlin, M.-W. Chang, K.~Lee, and K.~Toutanova, ``{BERT}: Pre-training of deep bidirectional transformers for language understanding,'' in \emph{Proceedings of the 2019 Conference of the North {A}merican Chapter of the Association for Computational Linguistics: Human Language Technologies, Volume 1 (Long and Short Papers)}.\hskip 1em plus 0.5em minus 0.4em\relax Association for Computational Linguistics, june 2019, pp. 4171--4186.

\bibitem{urvashi2019efficientlm}
U.~Khandelwal, K.~Clark, D.~Jurafsky, and L.~Kaiser, ``Sample efficient text summarization using a single pre-trained transformer,'' \emph{arXiv preprint arXiv:1905.08836}, 2019.

\bibitem{learning_to_summarize}
N.~Stiennon, L.~Ouyang, J.~Wu, D.~M. Ziegler, R.~Lowe, C.~Voss, A.~Radford, D.~Amodei, and P.~Christiano, ``Learning to summarize from human feedback,'' in \emph{Proceedings of the 34th International Conference on Neural Information Processing Systems}, ser. NIPS '20.\hskip 1em plus 0.5em minus 0.4em\relax Red Hook, NY, USA: Curran Associates Inc., 2020.

\bibitem{chaganty-etal-2018-price}
A.~Chaganty, S.~Mussmann, and P.~Liang, ``The price of debiasing automatic metrics in natural language evalaution,'' in \emph{Proceedings of the 56th Annual Meeting of the Association for Computational Linguistics (Volume 1: Long Papers)}, I.~Gurevych and Y.~Miyao, Eds.\hskip 1em plus 0.5em minus 0.4em\relax Melbourne, Australia: Association for Computational Linguistics, Jul. 2018, pp. 643--653.

\bibitem{ondrej2017nlg}
O.~Dusek, J.~Novikova, and V.~Rieser, ``Referenceless quality estimation for natural language generation,'' \emph{CoRR}, vol. abs/1708.01759, 2017.

\bibitem{ziegler2019fine}
D.~M. Ziegler, N.~Stiennon, J.~Wu, T.~B. Brown, A.~Radford, D.~Amodei, P.~Christiano, and G.~Irving, ``Fine-tuning language models from human preferences,'' \emph{arXiv preprint arXiv:1909.08593}, 2019.

\bibitem{ouyang2022training}
L.~Ouyang, J.~Wu, X.~Jiang, D.~Almeida, C.~Wainwright, P.~Mishkin, C.~Zhang, S.~Agarwal, K.~Slama, A.~Ray \emph{et~al.}, ``Training language models to follow instructions with human feedback,'' \emph{Advances in neural information processing systems}, vol.~35, pp. 27\,730--27\,744, 2022.

\bibitem{glaese2022improving}
A.~Glaese, N.~McAleese, M.~Trebacz, J.~Aslanides, V.~Firoiu, T.~Ewalds, M.~Rauh, L.~Weidinger, M.~Chadwick, P.~Thacker \emph{et~al.}, ``Improving alignment of dialogue agents via targeted human judgements,'' \emph{arXiv preprint arXiv:2209.14375}, 2022.

\bibitem{bai2022training}
Y.~Bai, A.~Jones, K.~Ndousse, A.~Askell, A.~Chen, N.~DasSarma, D.~Drain, S.~Fort, D.~Ganguli, T.~Henighan \emph{et~al.}, ``Training a helpful and harmless assistant with reinforcement learning from human feedback,'' \emph{arXiv preprint arXiv:2204.05862}, 2022.

\bibitem{ppo_2017}
\BIBentryALTinterwordspacing
J.~Schulman, F.~Wolski, P.~Dhariwal, A.~Radford, and O.~Klimov, ``Proximal policy optimization algorithms,'' \emph{CoRR}, vol. abs/1707.06347, 2017. [Online]. Available: \url{http://arxiv.org/abs/1707.06347}
\BIBentrySTDinterwordspacing

\bibitem{chowdhery2023palm}
A.~Chowdhery, S.~Narang, J.~Devlin, M.~Bosma, G.~Mishra, A.~Roberts, P.~Barham, H.~W. Chung, C.~Sutton, S.~Gehrmann \emph{et~al.}, ``Palm: Scaling language modeling with pathways,'' \emph{Journal of Machine Learning Research}, vol.~24, no. 240, pp. 1--113, 2023.

\bibitem{touvron2023llama}
H.~Touvron, T.~Lavril, G.~Izacard, X.~Martinet, M.-A. Lachaux, T.~Lacroix, B.~Rozi{\`e}re, N.~Goyal, E.~Hambro, F.~Azhar \emph{et~al.}, ``Llama: Open and efficient foundation language models,'' \emph{arXiv preprint arXiv:2302.13971}, 2023.

\bibitem{rafailov2024direct}
R.~Rafailov, A.~Sharma, E.~Mitchell, C.~D. Manning, S.~Ermon, and C.~Finn, ``Direct preference optimization: Your language model is secretly a reward model,'' \emph{Advances in Neural Information Processing Systems}, vol.~36, 2024.

\bibitem{soni-etal-2022-radqa}
S.~Soni, M.~Gudala, A.~Pajouhi, and K.~Roberts, ``{R}ad{QA}: A question answering dataset to improve comprehension of radiology reports,'' in \emph{Proceedings of the Thirteenth Language Resources and Evaluation Conference}, N.~Calzolari, F.~B{\'e}chet, P.~Blache, K.~Choukri, C.~Cieri, T.~Declerck, S.~Goggi, H.~Isahara, B.~Maegaard, J.~Mariani, H.~Mazo, J.~Odijk, and S.~Piperidis, Eds.\hskip 1em plus 0.5em minus 0.4em\relax Marseille, France: European Language Resources Association, Jun. 2022, pp. 6250--6259.

\bibitem{seo2016bidirectional}
M.~Seo, A.~Kembhavi, A.~Farhadi, and H.~Hajishirzi, ``Bidirectional attention flow for machine comprehension,'' in \emph{International Conference on Learning Representations}, 2016.

\bibitem{cui2017attention}
Y.~Cui, Z.~Chen, S.~Wei, S.~Wang, T.~Liu, and G.~Hu, ``Attention-over-attention neural networks for reading comprehension,'' in \emph{Proceedings of the 55th Annual Meeting of the Association for Computational Linguistics (Volume 1: Long Papers)}, 2017, pp. 593--602.

\bibitem{hu2019retrieve}
M.~Hu, Y.~Peng, Z.~Huang, and D.~Li, ``Retrieve, read, rerank: Towards end-to-end multi-document reading comprehension,'' in \emph{Proceedings of the 57th Annual Meeting of the Association for Computational Linguistics}, 2019, pp. 2285--2295.

\bibitem{liu2019roberta}
Y.~Liu, M.~Ott, N.~Goyal, J.~Du, M.~Joshi, D.~Chen, O.~Levy, M.~Lewis, L.~Zettlemoyer, and V.~Stoyanov, ``Roberta: A robustly optimized bert pretraining approach,'' \emph{arXiv preprint arXiv:1907.11692}, 2019.

\bibitem{yang2019xlnet}
Z.~Yang, Z.~Dai, Y.~Yang, J.~Carbonell, R.~R. Salakhutdinov, and Q.~V. Le, ``Xlnet: Generalized autoregressive pretraining for language understanding,'' \emph{Advances in neural information processing systems}, vol.~32, 2019.

\bibitem{chen-wu-2020-forcereader}
Z.~Chen and K.~Wu, ``{F}orce{R}eader: a {BERT}-based interactive machine reading comprehension model with attention separation,'' in \emph{Proceedings of the 28th International Conference on Computational Linguistics}, D.~Scott, N.~Bel, and C.~Zong, Eds.\hskip 1em plus 0.5em minus 0.4em\relax Barcelona, Spain (Online): International Committee on Computational Linguistics, Dec. 2020, pp. 2676--2686.

\bibitem{luo-etal-2020-map}
H.~Luo, Y.~Shi, M.~Gong, L.~Shou, and T.~Li, ``{M}a{P}: A matrix-based prediction approach to improve span extraction in machine reading comprehension,'' in \emph{Proceedings of the 1st Conference of the Asia-Pacific Chapter of the Association for Computational Linguistics and the 10th International Joint Conference on Natural Language Processing}, K.-F. Wong, K.~Knight, and H.~Wu, Eds.\hskip 1em plus 0.5em minus 0.4em\relax Suzhou, China: Association for Computational Linguistics, Dec. 2020, pp. 687--695.

\bibitem{raffel2020exploring}
C.~Raffel, N.~Shazeer, A.~Roberts, K.~Lee, S.~Narang, M.~Matena, Y.~Zhou, W.~Li, and P.~J. Liu, ``Exploring the limits of transfer learning with a unified text-to-text transformer,'' \emph{Journal of machine learning research}, vol.~21, no. 140, pp. 1--67, 2020.

\bibitem{ReasoNet2017}
\BIBentryALTinterwordspacing
Y.~Shen, P.-S. Huang, J.~Gao, and W.~Chen, ``Reasonet: Learning to stop reading in machine comprehension,'' in \emph{Proceedings of the 23rd ACM SIGKDD International Conference on Knowledge Discovery and Data Mining}, ser. KDD '17.\hskip 1em plus 0.5em minus 0.4em\relax New York, NY, USA: Association for Computing Machinery, 2017, p. 1047–1055. [Online]. Available: \url{https://doi.org/10.1145/3097983.3098177}
\BIBentrySTDinterwordspacing

\bibitem{hu2018reinforced}
M.~Hu, Y.~Peng, Z.~Huang, X.~Qiu, F.~Wei, and M.~Zhou, ``Reinforced mnemonic reader for machine reading comprehension,'' in \emph{Proceedings of the 27th International Joint Conference on Artificial Intelligence}, 2018, pp. 4099--4106.

\bibitem{lee2021mrc}
H.-G. Lee, Y.~Jang, and H.~Kim, ``Machine reading comprehension framework based on self-training for domain adaptation,'' \emph{IEEE Access}, vol.~9, pp. 21\,279--21\,285, 2021.

\bibitem{gharagozlou2022rlas}
H.~Gharagozlou, J.~Mohammadzadeh, A.~Bastanfard, S.~S. Ghidary \emph{et~al.}, ``Rlas-biabc: A reinforcement learning-based answer selection using the bert model boosted by an improved abc algorithm,'' \emph{Computational Intelligence and Neuroscience}, vol. 2022, 2022.

\bibitem{akrour2012april}
R.~Akrour, M.~Schoenauer, and M.~Sebag, ``April: Active preference learning-based reinforcement learning,'' in \emph{Machine Learning and Knowledge Discovery in Databases: European Conference, ECML PKDD 2012, Bristol, UK, September 24-28, 2012. Proceedings, Part II 23}.\hskip 1em plus 0.5em minus 0.4em\relax Springer, 2012, pp. 116--131.

\bibitem{chatgpt}
\BIBentryALTinterwordspacing
OpenAI, ``Chatgpt: Optimizing language models for dialogue,'' 2022. [Online]. Available: \url{https://openai.com/index/chatgpt/}
\BIBentrySTDinterwordspacing

\bibitem{chen-etal-2023-human}
J.~Chen, M.~Dodda, and D.~Yang, ``Human-in-the-loop abstractive dialogue summarization,'' in \emph{Findings of the Association for Computational Linguistics: ACL 2023}, A.~Rogers, J.~Boyd-Graber, and N.~Okazaki, Eds.\hskip 1em plus 0.5em minus 0.4em\relax Toronto, Canada: Association for Computational Linguistics, Jul. 2023, pp. 9176--9190.

\bibitem{nakano2021webgpt}
R.~Nakano, J.~Hilton, S.~Balaji, J.~Wu, L.~Ouyang, C.~Kim, C.~Hesse, S.~Jain, V.~Kosaraju, W.~Saunders \emph{et~al.}, ``Webgpt: Browser-assisted question-answering with human feedback,'' \emph{arXiv preprint arXiv:2112.09332}, 2021.

\bibitem{tunstall2023zephyr}
L.~Tunstall, E.~Beeching, N.~Lambert, N.~Rajani, K.~Rasul, Y.~Belkada, S.~Huang, L.~von Werra, C.~Fourrier, N.~Habib \emph{et~al.}, ``Zephyr: Direct distillation of lm alignment,'' \emph{arXiv preprint arXiv:2310.16944}, 2023.

\bibitem{zhao2023beyond}
Z.~Zhao, B.~Wang, L.~Ouyang, X.~Dong, J.~Wang, and C.~He, ``Beyond hallucinations: Enhancing lvlms through hallucination-aware direct preference optimization,'' \emph{arXiv preprint arXiv:2311.16839}, 2023.

\bibitem{bai2022constitutional}
Y.~Bai, S.~Kadavath, S.~Kundu, A.~Askell, J.~Kernion, A.~Jones, A.~Chen, A.~Goldie, A.~Mirhoseini, C.~McKinnon \emph{et~al.}, ``Constitutional ai: Harmlessness from ai feedback,'' \emph{arXiv preprint arXiv:2212.08073}, 2022.

\bibitem{Bradley1952RankAO}
R.~A. Bradley and M.~E. Terry, ``Rank analysis of incomplete block designs: I. the method of paired comparisons,'' \emph{Biometrika}, vol.~39, p. 324, 1952.

\bibitem{ramamurthy2023reinforcement}
R.~Ramamurthy, P.~Ammanabrolu, K.~Brantley, J.~Hessel, R.~Sifa, C.~Bauckhage, H.~Hajishirzi, and Y.~Choi, ``Is reinforcement learning (not) for natural language processing: Benchmarks, baselines, and building blocks for natural language policy optimization,'' 2023.

\bibitem{longpre2023flan}
S.~Longpre, L.~Hou, T.~Vu, A.~Webson, H.~W. Chung, Y.~Tay, D.~Zhou, Q.~V. Le, B.~Zoph, J.~Wei \emph{et~al.}, ``The flan collection: Designing data and methods for effective instruction tuning,'' in \emph{International Conference on Machine Learning}.\hskip 1em plus 0.5em minus 0.4em\relax PMLR, 2023, pp. 22\,631--22\,648.

\bibitem{johnson2016mimic}
A.~E. Johnson, T.~J. Pollard, L.~Shen, L.-w.~H. Lehman, M.~Feng, M.~Ghassemi, B.~Moody, P.~Szolovits, L.~Anthony~Celi, and R.~G. Mark, ``Mimic-iii, a freely accessible critical care database,'' \emph{Scientific data}, vol.~3, no.~1, pp. 1--9, 2016.

\bibitem{rajpurkar2016squad}
P.~Rajpurkar, J.~Zhang, K.~Lopyrev, and P.~Liang, ``Squad: 100,000+ questions for machine comprehension of text,'' in \emph{Proceedings of the 2016 Conference on Empirical Methods in Natural Language Processing}, 2016, pp. 2383--2392.

\bibitem{azar2024general}
M.~G. Azar, Z.~D. Guo, B.~Piot, R.~Munos, M.~Rowland, M.~Valko, and D.~Calandriello, ``A general theoretical paradigm to understand learning from human preferences,'' in \emph{International Conference on Artificial Intelligence and Statistics}.\hskip 1em plus 0.5em minus 0.4em\relax PMLR, 2024, pp. 4447--4455.

\bibitem{ethayarajh2024ktomodelalignmentprospect}
\BIBentryALTinterwordspacing
K.~Ethayarajh, W.~Xu, N.~Muennighoff, D.~Jurafsky, and D.~Kiela, ``Kto: Model alignment as prospect theoretic optimization,'' 2024. [Online]. Available: \url{https://arxiv.org/abs/2402.01306}
\BIBentrySTDinterwordspacing

\bibitem{liustatistical}
T.~Liu, Y.~Zhao, R.~Joshi, M.~Khalman, M.~Saleh, P.~J. Liu, and J.~Liu, ``Statistical rejection sampling improves preference optimization,'' in \emph{The Twelfth International Conference on Learning Representations}.

\end{thebibliography}
\bibliographystyle{IEEEtran}

\end{document}